\documentclass[12pt,preprint]{aastex}

\shorttitle{Hot Jupiter Frequencies from SuperLupus Survey} 
\shortauthors{Bayliss \& Sackett}

\begin{document} 

\title{The Frequency of Hot Jupiters in the Galaxy: \\Results from the
SuperLupus Survey} 

\author{Daniel D.\ R.\ Bayliss\altaffilmark{1}}
\email{daniel@mso.anu.edu.au}
\author{Penny D.\ Sackett\altaffilmark{1}}

 \altaffiltext{1}{Research School of Astronomy and
   Astrophysics, The Australian National University, Mt Stromlo
   Observatory, Cotter Rd, Weston Creek, ACT 2611, Australia}

\begin{abstract}
  We present the results of the SuperLupus Survey for transiting hot
  Jupiter planets, which monitored a single Galactic disk field
  spanning 0.66 deg${^2}$ for 108 nights over three years.  Ten
  candidates were detected: one is a transiting planet, two remain
  candidates, and seven have been subsequently identified as false
  positives.  We construct a new image quality metric, $S_j$, based on
  the behaviour of 26,859 light curves, which allows us to discard
  poor images in an objective and quantitative manner.  Furthermore,
  in some cases we are able to identify statistical false positives by
  analysing temporal correlations between $S_j$ and transit
  signatures.  We use Monte Carlo simulations to measure
  our detection efficiency by injecting artificial transits onto real
  light curves and applying identical selection criteria as used in
  our survey. We find at 90\% confidence level that
  0.10$^{+0.27}_{-0.08}$\% of dwarf stars host a hot Jupiter with a
  period of 1-10~days.  Our results are consistent with other transit
  surveys, but appear consistently lower than the hot Jupiter
  frequencies reported from radial velocity surveys, a difference we
  attribute, at least in part, to the difference in stellar
  populations probed.  In light of our determination of the frequency
  of hot Jupiters in Galactic field stars, previous null results for
  transiting planets in open cluster and globular cluster surveys no
  longer appear anomalously low.

\end{abstract}

\keywords{Planets and satellites: detection - Techniques: photometric}

\section{Introduction}
\label{sec:intro}

The discovery of short-period, giant extrasolar planets
\citep{1995Natur.378..355M, 1996ApJ...464L.147M} provided the exciting
potential for large
numbers of planets to be discovered by the transit method, as these
``hot Jupiters'' have a $\sim$10\% geometric probability of
transiting, and do so every few days.  However early
predictions greatly overestimated the actual discovery
rate \citep{2003ASPC..294..361H}.  This discrepancy resulted from
simplistic assumptions and a misunderstanding of the effects that
systematic noise would play in lowering detection efficiency
\citep{2006MNRAS.373..231P}.  It also resulted from an over-estimation
of the frequency of hot Jupiters.  More accurate, and lower,
predictions were provided in \citet{2008ApJ...686.1302B}, where it was noted
that objective and quantifiable detection criteria were required for
more robust inferences of planet frequencies.  By adopting such
objective and quantifiable detection criteria, we propose that the frequency of
hot Jupiters in the field has been overestimated by a factor of three \citep{2011OHP...11.01008}.

The SuperLupus Survey was established to detect hot Jupiters in a
field positioned just above the Galactic plane ($b=11^{\circ}$), and also to determine the
fraction of stars that host hot Jupiters in this typical Galactic
field.  The survey design and data analysis was constructed so as to
fulfill both of these objectives.  

In Section~\ref{observations} of this paper, we describe the SuperLupus
observations and data reduction.  The photometry is described in
Section~\ref{photometry}.  The criteria for candidate selection are
detailed in Section~\ref{sec:detection}, and the 10 identified candidates
are described and analyzed in Section~\ref{sec:candidates}.

In Section~\ref{mc_simulatinos}, we set out the details of the Monte
Carlo simulations performed to calculate the detection efficiency and
the effective number of stars probed for planets in the SuperLupus Survey.  In
Section~\ref{frequency}, we apply 
this efficiency to the actual results
of the survey to determine the fraction of stars in the
field that host a hot Jupiter.  Finally, in Section~\ref{discussion} we
summarize and discuss the implications of our results.

\section{Observations and Data Reduction}
\label{observations}

The SuperLupus Survey is an extension of the Lupus Survey
\citep{2009AJ....137.4368B}, in which the duration of the original survey was
approximately doubled to provide sensitivity to transiting planets
with periods as long as 10~days.  Many details of the data reduction procedure are
set out elsewhere \citep{2009AJ....137.4368B}; here we summarize
the procedure and highlight aspects that differed between the Lupus
and SuperLupus projects. 

Both surveys monitored a single field using the Wide-Field Imager on
the ANU 40-Inch Telescope at Siding Spring Observatory.  The field, in
the constellation of Lupus, is 11$^{\circ}$ above the Galactic plane
centered at R.A.$=$15$^{\rm{h}}$30$^{\rm{m}}$36.3$^{\rm{s}}$,
Decl.$=$$-$42$^{\circ}$53$'$53.0$\arcsec$ (J2000).  In total, 5158
images of 300~s exposure were taken in a custom ($V$+$R$) filter.
Of these, 2801 images were from the original Lupus Survey (2005 and 2006) and an
additional 2357 new images were taken in 2008.  The average full-width
half-maximum of the point spread function (PSF) over all 5158 images is
2.02\arcsec.

Initial data reduction (bias, dark, and flatfield correction) was
carried out on the mosaic frames using standard IRAF\footnote{IRAF is
  distributed by the National Optical Astronomy Observatories, which
  are operated by the Association of Universities for Research in
  Astronomy, Inc., under cooperative agreement with the National
  Science Foundation.} tasks in the package MSCRED.\, Sky flats were
obtained at twilight whenever conditions permitted, and master
flatfields were produced by median combining the $\sim$30 sky flats most
proximate in time to each night.  Bad pixels and columns were masked
and the mosaic frames were split into individual CCD frames.  A world
coordinate system solution was calculated for each image in order to
identify stars for photometry.

\section{Photometry and Detrending}
\label{photometry} 

Photometry for the original Lupus Survey was performed using Difference
Imaging Analysis
\citep[DIA:][]{1998ApJ...503..325A,2000AcA....50..421W}.  For the
SuperLupus Survey, we instead used Source Extractor
\cite[][]{1996A&AS..117..393B} aperture photometry.  This approach was
motivated by the fact that many of the survey images have PSFs that
are asymmetric due to tracking or focusing issues.  Such PSFs can
cause potential problems for DIA photometry, but are not of concern
for aperture photometry as long as the photometric aperture is large
enough.  Initial tests revealed that aperture photometry could achieve
the same (and even better) photometric precision for the majority of
the brighter stars ($V<20$) in the field.  In fact we note that the final
SuperLupus aperture photometry resulted in approximately 10$\%$ more stars
with a root-mean-squared (RMS) variability less than 0.025~mag compared with the original
Lupus Survey using DIA.

We found that a photometric aperture radius of 10 pixels (3.75\arcsec) resulted in the highest
precision lightcurves for the majority of these stars.

A reference catalog of 50,907 target
stars ($V<20$) was produced from a single, high quality, reference
image.  Aperture photometry was performed on these stars for each
survey image, with apertures re-centered on the point
of peak flux for each star in each image.  Background subtraction was performed
using a background map produced for each image by median filtering
and bi-cubic-spline interpolation over a 8$\times$8 pixel grid.

Systematic trends in the resulting lightcurves were removed using the
Sys-Rem algorithm \citep{2005MNRAS.356.1466T} with 12 iterations.  To further refine this dataset, we removed stars with lightcurves that
displayed RMS variability greater than 0.05~mag.
We also discarded poor quality images as described
below.  In total this left us with 26,859 stars over 3585 images.

\subsection{Image Quality Metric}

To identify poor quality images, we define an image quality metric ($S_{j}$) as: 
\begin{equation}
  \label{eq:sj}
  S_{j}^{2}=\sum_{i} \frac{r_{ij}^{2}}{\sigma^{2}_{ij}}~,
\end{equation}
where for the $i^{th}$ star on the $j^{th}$ image, $r_{ij}$ is the
average subtracted residual magnitude and $\sigma_{ij}$ is the
photometric uncertainty.  This metric is computed in the Sys-Rem
algorithm, where linear systematic effects are removed in order to
minimise $S_{j}$ \citep{2005MNRAS.356.1466T}.  In essence, $S_{j}$ is
a measure of how all the stars in image $j$ vary from the mean of all
images in the survey.  We retained only those images with $S_{j}<0.02$
for searching for planetary transits.  The advantage of using this
metric is that it uses information from all stars over the chip in a
single quality factor that is directly related to the precision of 
the resulting stellar photometry.  

This image quality metric also proved useful to determine if transit
candidates were statistical false positives by looking for 
correlations between ``transit'' events and $S_{j}$ (see Section~\ref{sec:candidates}). 

\section{Transit Search and Detection Criteria}
\label{sec:detection}
 
In order to accurately determine the efficiency of the SuperLupus
Survey, candidates must be identified in an automated and systematic
manner that can be applied identically to synthetic lightcurves in the
Monte Carlo simulation (see Section~\ref{mc_simulatinos}).  This was
implemented by way of a set of six selection criteria that were
automatically applied to each processed lightcurve.

Initially all 26,859 SuperLupus lightcurves were searched for transit
events using the Box-fitting Least-Squares (BLS) algorithm
\citep{2002A&A...391..369K}.  Trial periods in the range
$1.01<P<10$~d were tested with 55,000 equally-spaced frequency steps and 200 phase bins per trial
frequency.  Candidates were then identified based on their BLS Signal Detection
Efficiency ($SDE$), as defined in \citet{2002A&A...391..369K}.  After
testing synthetic lightcurves with the same temporal and noise characteristics as our data, 
a candidate threshold value of
$SDE=7.0$ was adopted. 

We defined the effective signal-to-noise of the
detected transit event, $\alpha$, for all candidates, as:
\begin{equation}
  \label{eq:alpha}
  \alpha=\frac{\delta}{\sigma}\sqrt{n_{obs}}~,
\end{equation}
where $\delta$ is the BLS-determined transit depth, $\sigma$ is the
RMS variability of the entire (Sys-Rem corrected) lightcurve, and
$n_{obs}$ is the actual number of data points within the
BLS-determined transit event, i.e. the number of ``in-transit'' data
points.  A threshold of $\alpha=10$ was adopted, identical to the
threshold adopted in two other transit surveys of faint stars
\citep{2006AJ....132..210B,2009ApJ...695..336H}.  At this stage we
also rejected candidates for which more than 90\% of
in-transit data originated from an event on a single night.

As with other ground-based surveys, many of our candidates displayed
integer-day periods resulting from nightly systematic effects.
Therefore candidates with $1.01<P<1.02$ and $1.98<P<2.02$ were
rejected.  We also required that candidates had a magnitude of
$V<18.8$.

Finally candidates were cross-matched with the catalog of known
variable stars in the field \citep{2008AJ....135..649W} and eclipsing
binary systems were removed.

The selection criteria are summarized in Table~\ref{tab:criteria}.

\section{Initial Candidates}
\label{sec:candidates}

A total of 10 lightcurves fulfilled the six selection criteria set out
in Section~\ref{sec:detection}.  Their characteristics are described in
Table~\ref{tab:candidates}.

The candidate SL-07 was immediately recognised as Lupus-TR-3, a star
with a transiting planet that had been discovered during the original
Lupus Survey \citep{2008ApJ...675L..37W,2009AJ....137.4368B}.
To investigate the likely nature of the other candidates, we used
the $\eta_{p}$ diagnostic method proposed by
\citet{2005ApJ...627.1011T}.  The results are set out in Table~\ref{tab:candidates2}.  Candidates SL-02 and SL-04 show very high diagnostic numbers, 
indicating that their transit parameters make them unlikely to be real
transiting planets.  In the case of candidate SL-04, a very
clear transit was observed on 6 June 2005, with a transit time of approximately 6~hr, in contrast
to the BLS determined duration of 8.08~hr.  Use of the shorter
duration reduces the $\eta_{p}$ diagnostic value to 1.9, which, while still high, could be
explained by a large radius planet, since the $\eta_{p}$ diagnostic is
normalized to the assumption $R_{P}=1R_{J}$.

To help identify statistical false positives, we checked if the
detected transit event was correlated with the image quality, using the calculated value of the metric $S_{j}$, as
defined in Eq.~(\ref{eq:sj}).

\begin{figure}[htbp]
  \centering
  \plotone{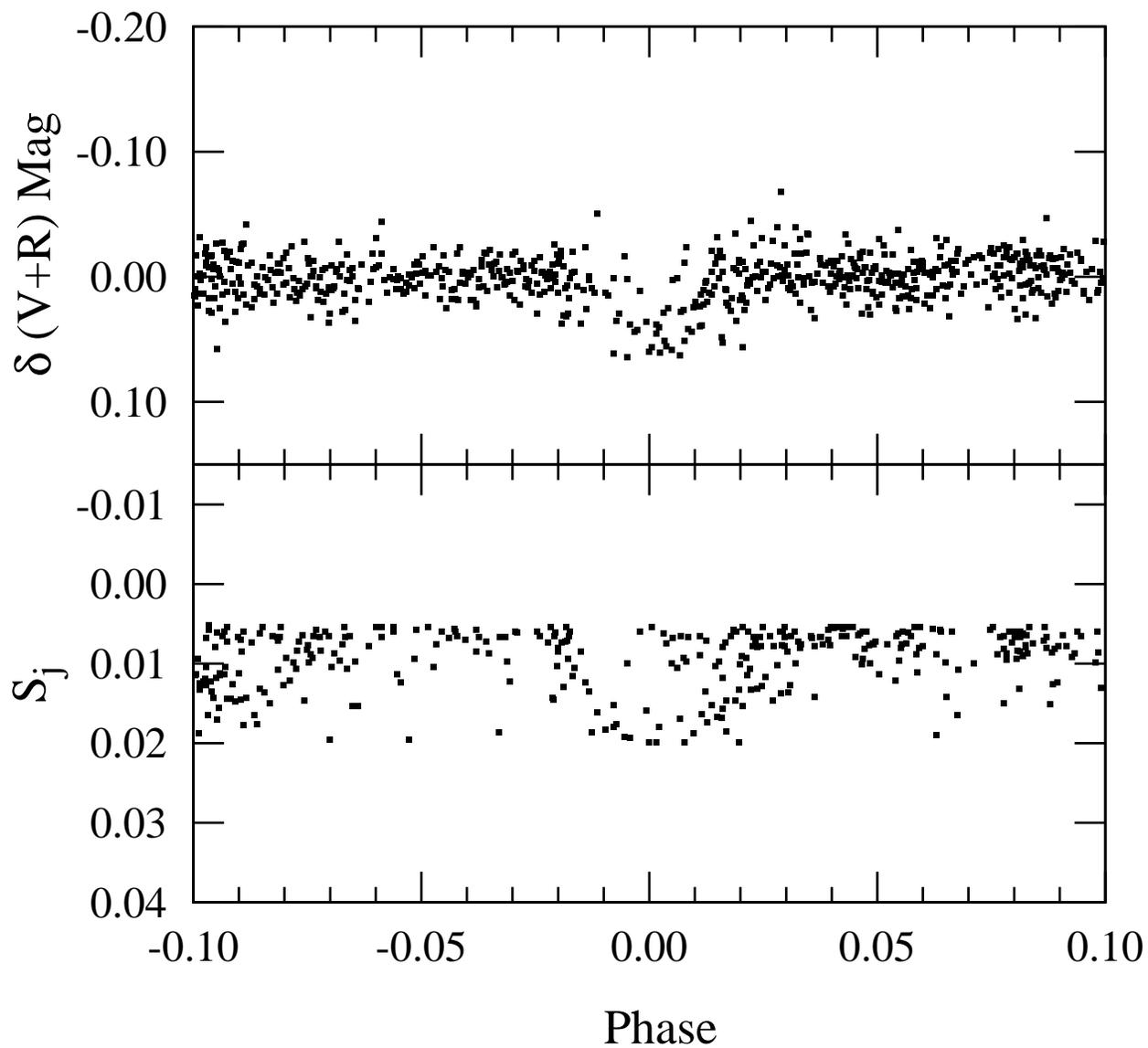}
  \caption{Phase-wrapped lightcurve for candidate SL-08
    (top panel) and the corresponding image quality, $S_{j}$, for
    images with $S_{j}>0.005$ (bottom panel) wrapped at the same
    period and phase.  The correlation between the transit event and
    image quality indicates that SL-08 is a statistical false positive.}
  \label{fig:correlation}
\end{figure}

Three candidates, SL-08, SL-09 and SL-10, were found to have transit events that
were strongly correlated with $S_{j}$, indicating that the detected transit
signature is almost certainly a systematic effect that the Sys-Rem algorithm
had not fully removed.  An example of such a correlation is shown in
Figure~\ref{fig:correlation}. 

We inspected the images of the candidates for signs that the
transit signature might be caused by proximity to the edge of a CCD, a bad
column, or a bright star.  We also inspected the image for evidence of
significant flux contributions from neighboring stars within the photometric aperture.

SL-01 was found to be blended with a very bright, saturated
neighboring star at a distance of 27\arcsec.  We therefore
ruled this out as being a genuine candidate.  We found that the candidate SL-03 had three neighbors close to the
photometric aperture, and that SL-05 had a neighbor well within the
photometric aperture.  Photometry for these candidates was re-derived
using a smaller photometric aperture to avoid the contaminating
neighbors.  This analysis revealed that the transit event seen in
SL-05 was in fact due to a very deep eclipse (40\%) occurring in a
faint neighbour situated just $2.6\arcsec$ from SL-05.  It also
revealed that while the SL-03 target star was responsible for the observed transit
event, close neighbors and the large aperture of the original
photometry had diluted the size of the dip from the true
depth of 0.100~mag to a shallower 0.042~mag.  
The true depth is characteristic of an eclipsing binary and does not 
pass our selection criterion.

In summary, we conclude that of the 10 candidates identified in
the SuperLupus Survey, only SL-07 can be confirmed as a genuine
transiting planet.  We rule out seven candidates (SL-01, SL-02, SL-03,
SL-05, SL-08, SL-09, SL-10) as being false
positives, while two candidates (SL-04 and SL-06) remain.  These two
remaining candidates are V=17.8 and V=18.6 respectively, making
follow-up extremely difficult.  
Consequently, we have not observed these candidates further, 
although formally we cannot exclude them.  In
Section~\ref{frequency} we calculate hot
Jupiter frequencies assuming only SL-07 is a transiting
planet, but also provide figures for the case where one of these
remaining candidates is a transiting planet.

\section{Monte Carlo Simulations to Determine Detection Efficiency}
\label{mc_simulatinos}

To determine the efficiency of the SuperLupus Survey to detecting
transiting planets, we carried out a Monte Carlo simulation of the
SuperLupus Survey.  Synthetic transit signatures were inserted into 
actual SuperLupus lightcurves
and then recovery attempted using the same methodology and selection
criteria used for our survey.

\subsection{Modeling the Stellar Population}
\label{sec:besancon}

\begin{figure}[htbp]
  \epsscale{0.7}
  \plotone{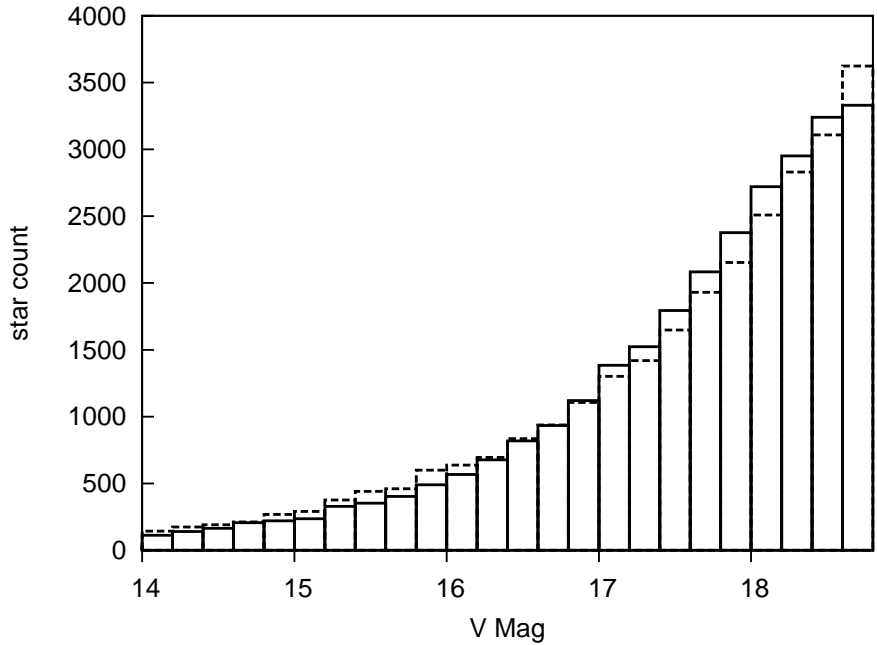}
  \caption{Comparison of the apparent V magnitude distribution between the
    Beason\c{c}on model (dashed line) and our observed field (solid line).}  
  \label{fig:beasoncon}
\end{figure}

Generating a realistic set of synthetic transit lightcurves against
which to test a detection algorithm requires knowledge of the actual
distribution of transit depths, durations, and periods of transiting
planets.  Ideally, this would be achieved by determining the radius of
each monitored star in the field.  This is possible when monitoring an
equidistant population such as an open cluster or a globular cluster
\citep[e.g.][]{2009ApJ...695..336H,2005ApJ...620.1043W}.  Since the
distances to field stars cannot be determined from photometry alone,
the exact radius of each star in the SuperLupus field is unknown.  We
therefore used the Beason\c{c}on model of the Galaxy
\citep{2003A&A...409..523R} to provide a statistical distribution of
stellar radii and masses for the stars monitored in our field.  The
population was synthesized over a 0.66~deg$^{2}$ field centered at
$b=11^{\circ}, l=331^{\circ}$.  Interstellar extinction was set at
0.6~mag\,kpc$^{-1}$, based on the Schlegel dust maps of the region
\citep{1998ApJ...500..525S}.  Stars were selected in the magnitude
range $14<V<18.8$ to match the SuperLupus observational and selection
constraints.  The Beason\c{c}on model returned a population of stars
that closely matched both the total star count, and the distribution of
apparent magnitudes (see Figure~\ref{fig:beasoncon}), giving us
confidence this model accurately simulates our field.

We note that the Beason\c{c}on model indicated
that 24\% of the survey stars have $\log{g}<4.0$.  In our Monte Carlo
these stars are classified as giants and given zero probability for
transit detection.  We assign a radius to each star in the simulation
based on the stellar mass from the Beasan\c{c}on model and the simple
relationship given by \citet{2000asqu.book.....C}:

\begin{equation}
  \label{eq:mass-radius}
  \log\frac{R}{R_{\odot}}=0.917\log\frac{M}{M_{\odot}}-0.020~.
\end{equation}

\subsection{Lightcurve Generation}
\label{sec:generation}

Lightcurves for the Monte Carlo were created by first randomly selecting a star from the
Beasan\c{c}on model population.  The temporal sampling of the star is
set to the actual timestamps of the SuperLupus Survey.  We then need
to attach an appropriate photometric uncertainty to each point on the
synthetic curve.  

This was done by using the $V$ magnitude of the
chosen star to randomly select an RMS uncertainty from the 
actual distributions in RMS observed in the SuperLupus
Survey for stars of that $V$ magnitude, as described below.

\begin{figure}[htbp]
  \epsscale{0.8}
  \plotone{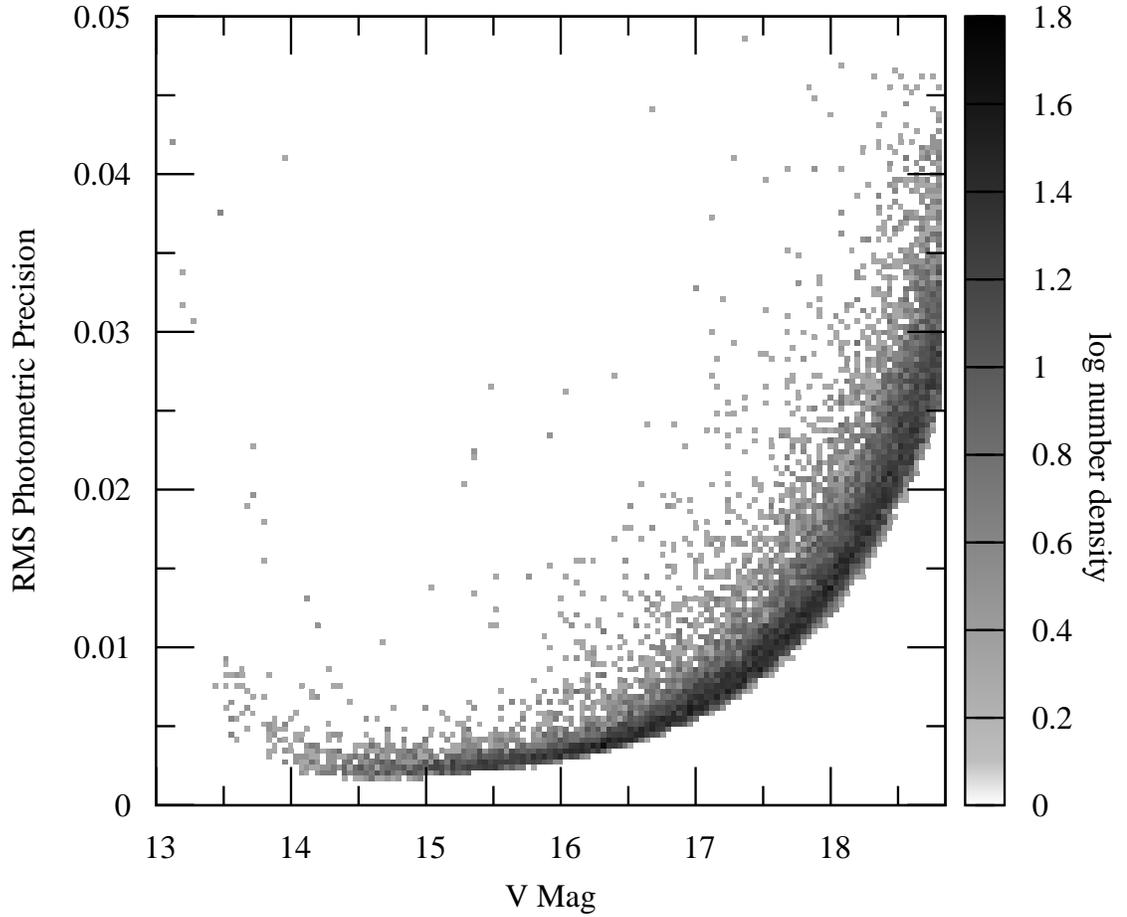}
  \caption{The number density of stars with a given RMS photometric precision
  plotted against $V$ mag.  The number density is given on a
  logarithmic scale.  The bin width is 0.04~mag for $V$ magnitude, and
  0.3 mmag for RMS photometric precision.}
  \label{fig:rms-density}
\end{figure}

\begin{figure}[htbp]
  \epsscale{0.8}
  \plotone{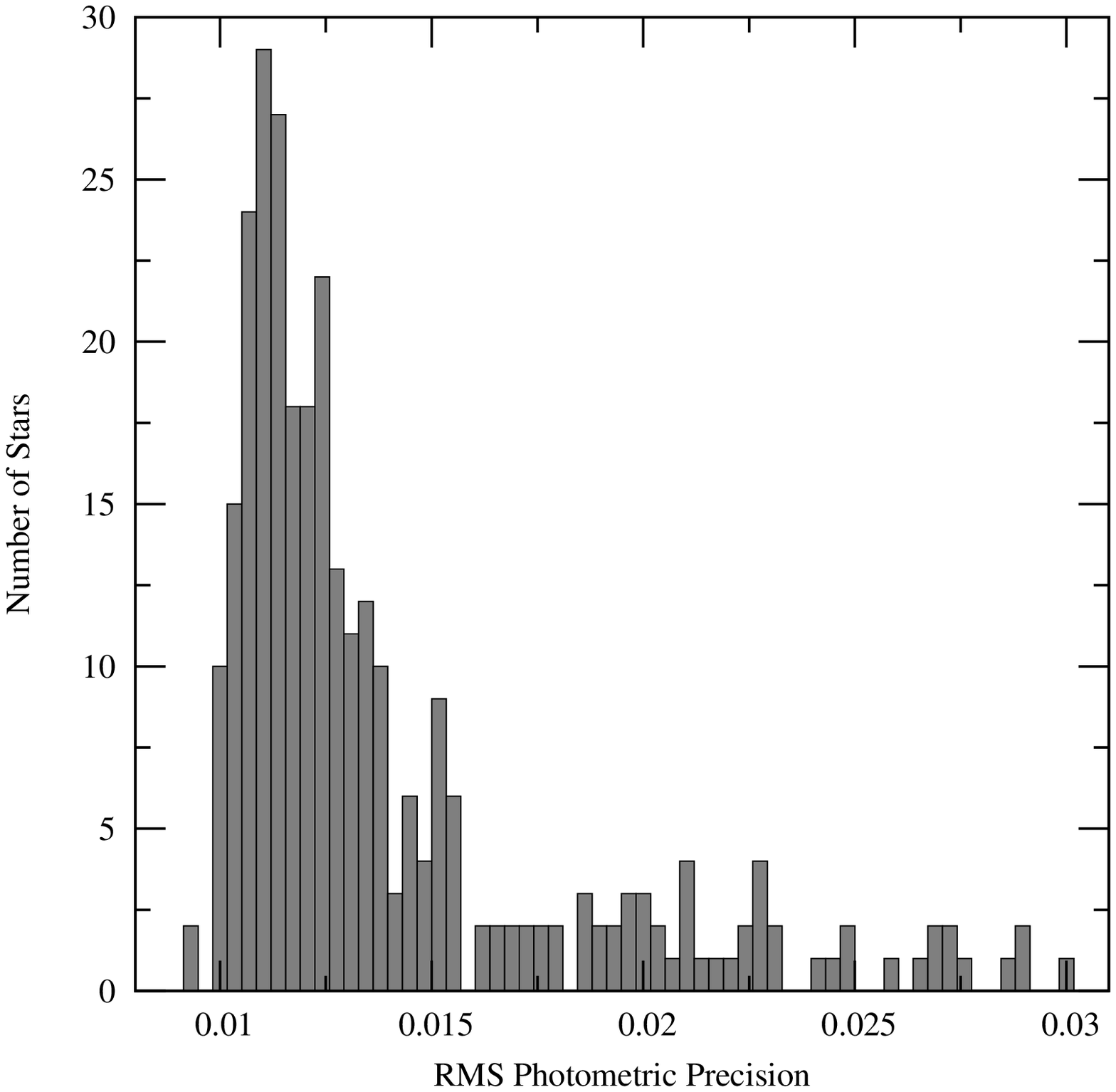}
  \caption{RMS photometric precision distribution profile for stars in
    the SuperLupus Survey between $V=17.70$ and $V=17.74$. }  
  \label{fig:rms_scatter_dist}
\end{figure}

The RMS scatter for each real lightcurve in the SuperLupus Survey,
after the application of Sys-Rem, shows a wide
range in RMS scatter at a given $V$ magnitude.  Therefore we
calculated the actual distribution of RMS scatter for each $V$~magnitude in
the survey, as displayed in Figure~\ref{fig:rms-density}.  As an example, the distribution for $V=17.72$ is shown in
Figure~\ref{fig:rms_scatter_dist}. It can be thought of as
a vertical slice through the density plot in
Figure~\ref{fig:rms-density}.

It is evident that the RMS scatter distributions are non-Gaussian,
with long tails towards high RMS scatter.  This further highlights the
importance of basing the RMS scatter of synthetic lightcurves on the
actual distributions of photometric precision rather than
approximating them using a single-valued function of magnitude as has
been done in some previous studies \citep[e.g.][]{2005ApJ...620.1043W,2006AcA....56....1G}.

A value for the photometric precision is selected randomly, using the
appropriate RMS scatter distribution as a probability weighting, for each
Beason\c{c}on star.  Once the value of the photometric precision is
determined, a real SuperLupus lightcurve is selected with the same
photometric precision, and used as the base lightcurve into which we
inject a transit.

\subsection{Injecting Transits}
\label{sec:injecting}

To simulate the lightcurve from a transiting hot Jupiter, we first
determined the parameters for each simulated system.  The stellar mass
and radius were taken from the Beason\c{c}on model and Eq.~\ref{eq:mass-radius}.  The planetary
mass was fixed at $1~M_{J}$, but we note that this does not
significantly alter the transit parameters when $M_{P}<<M_{\star}$.
Four different planetary radii were simulated: $R_{P}=$ 0.8, 1.0, 1.2,
and 1.4~$R_{J}$.

The inclination of the orbit, $i$, was determined by choosing a random
$\cos i$ between zero and $\cos i_{min}$, the
limiting inclination that results in a transit \citep{1999poss.conf..189S}:

\begin{equation}
  \label{eq:cosi}
  \cos i_{min}=\frac{R_{\star}+R_{P}}{a}~,
\end{equation}

where $a$ is the planet's semi-major orbital axis.

Although there is some evidence for a ``pile-up'' of planets at
$\sim$3~day periods \citep{2008PASP..120..531C}, the intrinsic
distribution of planets with periods between 1-10~days is not well
understood.  Therefore it is necessary to make an assumption in
assigning the period distribution in the Monte Carlo simulation.
\citet{2009ApJ...695..336H} use a uniform logarithmic distribution.  We test this distribution, as well as a uniform linear
distribution in period.

With the parameters $M_{\star}$, $M_{P}$, $R_{\star}$, $R_{P}$, $\cos
i$, and $P$ assigned to each star, a transit depth and duration was
calculated and a box-shaped transit inserted into each synthetic
lightcurve.  For each parameter set, we simulated 10 randomly-selected
phases.  We adopted a simple box-shape model as limb darkening effects
have been shown to be negligible in this signal-to-noise regime
\citep{2006AcA....56....1G}.  The transits were inserted into
lightcurves that had already been detrended using Sys-Rem, whereas a
transit signal in our survey would need to pass through the SyS-Rem
detrending.  We therefore carefully tested and optimised the number of
Sys-Rem iterations we used so that we would not degrade a real
astrophysical signal.  Only those trends that are present in many
light curves with the same temporal timestamps are removed.

\subsection{Detection Efficiency and Effective Number of Stars Probed}
\label{sec:detectionefficiency}

The BLS algorithm \citep{2002A&A...391..369K} was applied to each synthetic lightcurve, using the
same parameters used for the actual SuperLupus
dataset (see Section~\ref{sec:detection}). 
In total, 800,000 lightcurves were generated for the
Monte Carlo and searched
for transit signatures.  
Transits were deemed to be detected if the selection criteria 
used in the actual SuperLupus search (see Section~\ref{sec:detection}) were satisfied.

The detection efficiency, $\varepsilon$, is defined as the
fraction of lightcurves in which the 
inserted transit signal is detected in the Monte Carlo simulation:
\begin{equation}
  \label{eq:DE}
  \varepsilon=\frac{1}{N_{sim}}\sum^{N_{sim}}_{i=1} \delta_{i,}~,
\end{equation}
where $\delta_{i}$ is a delta function representing whether the transit of star $i$ is
detected ($\delta_{i}=1$) or not detected ($\delta_{i}=0$), and
$N_{sim}$ is the number of simulated stars in the Monte Carlo
simulation.  As discussed in Section~\ref{sec:besancon}, 24$\%$ of our survey
stars are giants ($\log (g)<4.0$), and were automatically assigned a
$\delta_{i}=0$ regardless of the other system parameters.  We also
note that only transiting systems were simulated, so $\varepsilon$
does not include the geometric probability of transit.

To determine the frequency of hot Jupiters in the field of the
SuperLupus Survey we need to determine how many dwarf stars we have effectively probed for transiting planets,
$N_{pr}$.  This number depends on the geometric probability of
transit, the number of stars monitored to a given precision (for SuperLupus this is
20,465), as well as the detection efficiency.  We calculated the
geometric probability of transit, $f_{i}^{geo}$, for each system generated in our Monte Carlo
simulation. $N_{pr}$ is therefore given as: 

\begin{equation}
  \label{eq:n_probed}
  N_{pr} = \frac{N_{sur}}{N_{sim}}\sum^{N_{sim}}_{i=1}
  \delta_{i} f^{geo}_{i} ~.
\end{equation}

Both the detection efficiency and the effective number of dwarf stars probed are set
out in Table~\ref{tab:montecarlo} for a variety of period ranges and
assumed period distributions of hot Jupiters.  Also
tabulated are the mean transit depth, transit duration, and
mean transit probability for each set of Monte Carlo assumptions about
the size and period distribution of hot Jupiters.  The results are presented for all periods
simulated (1-10~days), as well as for period
ranges of 1-3~days, 3-5~days, and 5-10~days in order to allow for direct
comparison with the results of \citet{2006AcA....56....1G} and
\citet{2009ApJ...695..336H}. 

\begin{figure}[htbp]
  \centering
  \plotone{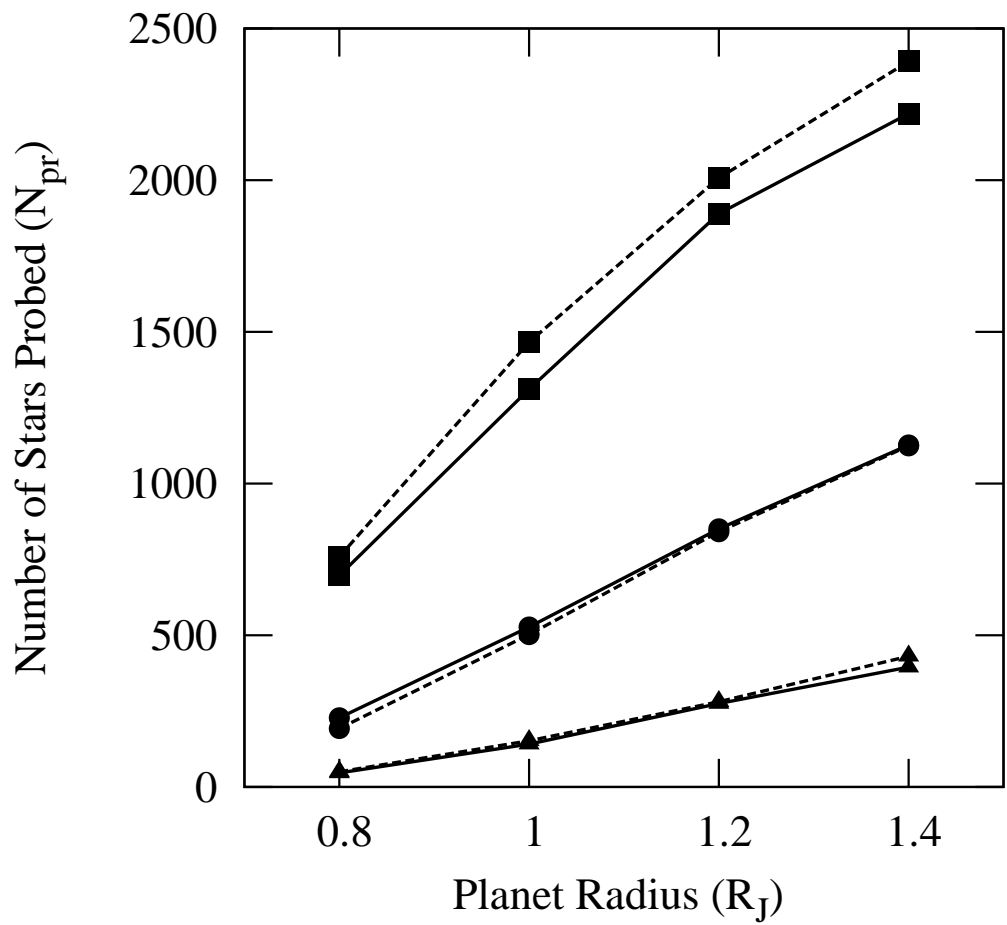}
  \caption{The effective number of dwarf stars probed ($N_{pr}$) for transiting hot Jupiters in
    the SuperLupus Survey as a function of planet
    radius.  The solid lines represent the assumption of a uniform period
    distribution, while the dashed lines represent the assumption of a log uniform
    distribution.  The square symbols are for 1-3~d periods, the
    circles for 3-5~d periods, and the triangles for 5-10~d periods.}
  \label{fig:survey_eff} 
\end{figure}

In Figure~\ref{fig:survey_eff}, $N_{pr}$ is plotted as a function of
planet radii for each of the four planetary radii simulated.  As
expected, more stars are probed in all period ranges for large radii
planets, and more stars are probed for shorter period planets.  The
difference between the uniform and uniform logarithmic period
distributions is most marked for the 1-3~d period planets.

\section{Frequency of Planets in the Survey}
\label{frequency}

The frequency of dwarf stars that host hot Jupiter planets, $f$, can be
calculated simply
using the Monte Carlo results and the SuperLupus Survey results:
\begin{equation}
  \label{eq:freq_HJ}
  f = \frac{n}{N_{pr}}~,
\end{equation}
where $n$ is the number of planets detected in the survey.

With only one (or possibly two) planets detected, Poisson statistics
are used to determine confidence intervals, or upper limits in the
cases of no planet detections.  The probability of detecting $n$ planets
when $\lambda$ planets are expected is:
\begin{equation}
  \label{eq:poisson}
  P(n,\lambda)=e^{\lambda}\frac{\lambda^{n}}{n!}~.
\end{equation} 
Since the underlying distribution of planets is unknown, a Bayesian
approach is used, with a flat prior for the uniform distribution, and
a log prior for the logarithmic distribution.  The upper and lower
90\% confidence intervals ($\sigma_{high}$ and $\sigma_{low}$,
respectively) are calculated by solving:
\begin{equation}
  \label{eq:poisson_twoside}
  \frac{\int^{\sigma_{high}}_{\sigma_{low}}\frac{e^{\lambda}\lambda^{n}}{n!}d\lambda}
  {\int^{\infty}_{0}\frac{e^{\lambda}\lambda^{n}}{n!} d\lambda     }= 0.9~.
\end{equation}
The values of $\sigma_{high}$ and $\sigma_{low}$ were determined numerically from
Eq.~\ref{eq:poisson_twoside}, with a symmetric confidence interval such
that 5\% of the probability distribution was below $\sigma_{low}$ and
5\% above $\sigma_{high}$.  The limits $\sigma_{high}$ and $\sigma_{low}$
were then used to calculate the upper and low limits for 90\%
confidence intervals.

Where no planets were detected in a period range, a one-sided 95\% confidence
upper limit is used so that it can be directly compared with the two-sided
90\% limits.  In that case, $n=0$, and Eq.~\ref{eq:poisson_twoside}
simplifies to: 
\begin{equation}
  \label{eq:poisson_oneside}
  1-e^{-\sigma_{high}} = 0.95~.
\end{equation}
This method of calculating confidence limits is similar to that used
by \citet{2006AcA....56....1G} and \citet{2009ApJ...695..336H}.

The follow-up work presented in \citet{2008ApJ...675L..37W} reveals that one of the SuperLupus candidates, SL-07, is indeed a hot Jupiter planet.
Seven other candidates can be ruled out as false positives,
while SL-04 and SL-06 remain as candidates, although follow-up would
be difficult due to their faintness.  We therefore
present our frequency results for the case where Lupus-TR-3b is the
only transiting planet detected ($n=1$) and also for the case where
either SL-04 or SL-06 is also a transiting planet ($n=2$).  The
results are set out in Table~\ref{tab:freq_SL}, again for both uniform and uniform
logrithmic period distributions, and are tabulated in the
same period ranges as used in Table~\ref{tab:montecarlo}.  Frequencies
for each of the four simulated planet radii are calculated, along with
the mean over these radii, which is $R_{P}=1.1$.

\subsection{Comparison with Other Surveys}
\label{sec:comparison}

There are four other transit studies of non-cluster fields that report statistics for
hot Jupiter frequency.

The {\it Kepler Mission\/} \citep{1997ASPC..119..153B} yields very high
precision, near continiuos photometry for around 150,000 target stars
in a selected galactic field.  \citet{2011arXiv1103.2541H} calculate the the frequency of planets within
0.25AU of solar-type hosts stars selected from the Kepler target
stars.  For the period range of 1.2$<$P$<$10~d, and radii
0.7$<R_{J}<$1.4, very similiar to those considered in this work, the frequency of hot Jupiters is given as
0.37$\%$ \citep[Figure~4 in][]{2011arXiv1103.2541H}.  This frequency is based only on ``solar type'' Kepler target stars, 
defined as those with temperatures of $T_{eff}=4100-6100$~K and gravities of
$4.0<\log{g}<4.9$.  While this gravity cut is essentially the same as
is used in this work, the temperature cut means that a more limited
sample of stars are being probed.  Coupled with the fact that these
stars are already drawn from a set of targets selected using
``detectability metrics'' that suggested they were most likely to give
a detectable transit signal for terrestrial-size planets \citep{2010ApJ...713L.109B}, the frequency
of 0.37$\%$ is higher than one might expect for ground based survey of
random Galactic dwarfs in the field.  
Indeed the Kepler result is at the 
high end of the frequency of hot Jupiters determined from our SuperLupus survey, 
just consistent within our uncertainty.

The MMT transit survey of the open cluster M37 \citep{2009ApJ...695..336H} included a significant number
of non-cluster field stars, and these were analyzed separately to
determine the hot Jupiter frequency for Galactic dwarf stars.  Assuming a
Jupiter-sized planet, and a logarithmic period distribution, that study found that the fraction of field stars
hosting planets with periods of 1-3~d is $<$0.8\%, and those with
planets having periods of 3-5~d is $<$2.7\%.  These statistics are
upper limits only, as there were no confirmed transiting planets.

Analysis of the OGLE-III survey for transiting planets puts the
frequency of hot Jupiters at 0.14\% for 1-3~d period planets, and
0.31\% for periods of 3-5~d \citep{2006AcA....56....1G}.  These
statistics draw on survey fields in both the Galactic plane and the
Galactic bulge, and were derived based on five transiting planets that
had been confirmed from the OGLE-III survey.  They rely on an
estimation of the efficiency of the ``by-eye'' selection that was used
by the OGLE team to select transiting planet candidates. 

The Hubble Space Telescope (HST) was used to undertake a survey for
transiting planets in the crowded bulge region of the Galaxy
\citep{2006Natur.443..534S}.  Sixteen candidates were discovered
from a sample of 180,000 stars monitored, however only one of these
(SWEEPS-11) was bright enough to be confirmed in the usual manner
using radial velocity measurements.  Assuming that all the candidates are
indeed planets, it was reported that 0.42\% of bulge stars
more massive than $0.44~M_{\odot}$ have giant planets with periods
up to 4.2~days \citep{2006Natur.443..534S}.

Our result for 1-3~d period planets, $f<0.13\%$, is consistent with
the limits from the deep MMT survey of \citet{2009ApJ...695..336H}.
It is lower than the OGLE results, but consistent within the lower
half of their 90\% confidence interval.  For 3-5~d period planets, our
result of 
$f=0.15^{+0.41}_{-0.11}$ is consistent with the upper limit from the deep
MMT survey.  It is lower than the OGLE results, but still
within the 90\% confidence intervals.  For 5-10~d periods we are able
to place limits on the frequency of hot Jupiters
of $f<0.93\%$, assuming a logarithmic distribution in period and also
that SL-04 and SL-06 do not host genuine hot Jupiters.  If one of
these candidates is a transiting hot Jupiter this frequency in this
period range would be $0.44^{+1.19}_{-0.34}\%$, closer to the Kepler
figure of 0.37\% but still lower than the RV survey results.  The comparisons are summarised in
Table~\ref{tab:fraction}.

The frequency of hot Jupiters can also be determined from radial velocity
surveys.  \citet{2008PASP..120..531C} report a frequency of $0.65\pm
0.4\%$ from the Keck Planet Search, while
\citet{2005ESASP.560..833N} report a frequency of $0.7 \pm 0.5\%$
from the ELODIE planet search.  These comparisons are also set out in
Table~\ref{tab:fraction}.  Both of these radial velocity results are above our upper
90\% confidence interval (0.56\%), although the confidence intervals 
of the two survey results overlap.  

There are, however, major differences between transit surveys
and these radial velocity surveys that could lead to discrepancies in the frequency
derived for hot Jupiters.  Radial velocity surveys are sensitive to planetary
mass, while transit surveys are sensitive to planetary radius.  Also transit and radial velocity surveys usually
monitor different stellar
populations.  Radial velocity surveys typically monitor carefully
selected bright, sun-like stars.  More recently, bright M dwarfs have also
been monitored \citep{2009A&A...493..645F}.  Transit searches monitor
every star in the survey field that has the requisite signal-to-noise
to yield high precision photometry.  In a deep survey such as the
SuperLupus Survey, this will translate to a stellar population with a
sub-solar mean mass (0.9$M_{\odot}$ in the case of the SuperLupus
Survey).  If hot Jupiters of a given period are less frequent around lower mass stars,
then the frequency of hot Jupiters from deep
transit surveys will be lower than radial velocity surveys.

\section{Discussion}
\label{discussion}

The frequency of hot Jupiters in the Galaxy is an important quantity
that will ultimately provide a constraint on models of planet
formation and migration.

It has been suggested that the frequency of hot Jupiters in globular
cluster environments is lower than that of field stars, and that this
may be due to crowding or low metallicity affecting planet formation,
migration, or survival \citep{2000ApJ...545L..47G,2005ApJ...620.1043W}.  
Our results indicate, however, that there is little
statistical disagreement between hot Jupiter frequencies in cluster
and non-cluster environments, even if one of the remaining candidates
turns out to be a genuine planet.  This is consistent with the work
presented in \citep{2011...arXiv:1009.3013v1} which concludes that
there is no evidence to support that open clusters have a lower
frequency of hot Jupiters.

Initial estimates for planet yields from transit surveys turned out to
be far in excess of the actual discovery rate
\citep{2003ASPC..294..361H}.  One of the many factors that led to
this over-estimation was the adoption of the hot Jupiter frequency
derived from early radial velocity surveys, which as we have shown is higher than
is found from transit surveys.  
The {\it Kepler\/} result of 0.37$\%$
\citep{2011arXiv1103.2541H} obviously provides a robust statistic for
hot Jupiter frequencies due to the high detection efficiency of that
survey.  However the result should be adopted cautiously when
calculating expected yields from typical transit surveys, as it is based on a
sample of ``solar type'' stars drawn from Kepler target stars, rather
than the ensemble field stars monitored by most blind transit surveys.

\acknowledgments The authors thank David Weldrake and Brandon
Tingley, who 
initiated the original Lupus Survey with PDS.  
We also thank Grant Kennedy and Tom Evans for
assisting in gathering data for the SuperLupus project.

{\it Facilities:} \facility{SSO:1m (WFI)}

\clearpage

\bibliographystyle{apj}

\clearpage

\begin{deluxetable}{lc}
\tablewidth{0pt}
\tablecaption{Candidate selection criteria \label{tab:criteria}}
\tablehead{
\colhead{Criteria} & \colhead{Threshold}
}
\startdata
BLS Signal Detection Efficiency & $SDE > 7.0$ \\
Effective S/N of transit & $\alpha > 10$ \\
BLS-Period & $ P\not\le 1.02$~d \\
 & $P \ne 2.00 \pm 0.02$~d  \\
Magnitude & $V<18.8$ \\
Single-event fraction & $frac. < 0.9$ \\
Not a variable star & \citet{2008AJ....135..649W}\\
 & catalog
\enddata
\end{deluxetable}

\begin{deluxetable}{lcccccccccc}
\rotate
\tablewidth{0pt}
\tablecaption{Properties of Initial SuperLupus Candidates\label{tab:candidates}}
\tablehead{ 
\colhead{ID}&\colhead{R.A.}&\colhead{Decl.}&\colhead{Mag}&\colhead{Period}&
\colhead{Depth}&\colhead{Duration}&\colhead{$T_{C}$}&
\colhead{$SDE$}&\colhead{$\alpha$}&\colhead{$\eta_{p}$} \\
  & \colhead{(J2000)}&\colhead{(J2000)} & \colhead{($V$)} &\colhead{(days)}&\colhead{($V$+$R$)} &
\colhead{(hr)}&\colhead{(HJD)}&&&
}
\startdata
    SL-01  & 15:30:47.11 & -42:44:29.5  &  16.8 & 8.59642 & 0.006 &6.03  &2454616.0322 &  7.53 & 11.4  & 1.2  \\   
    SL-02  & 15:31:15.28 & -42:47:28.4  &  16.7 & 7.44930 & 0.005 &10.32 &2453894.0654 &  8.00 & 13.1  & 2.1  \\   
    SL-03  & 15:30:43.66 & -43:01:43.8  &  18.2 & 5.38421 & 0.042 & 3.66 &2453880.1516 &  8.60 & 21.0  & 1.4  \\   
    SL-04  & 15:31:37.13 & -43:10:16.9  &  17.8 & 7.72108 & 0.036 & 8.08 &2453883.2487 &  8.30 & 32.4  & 1.9  \\   
    SL-05  & 15:30:24.24 & -43:15:10.7  &  18.4 & 1.88490 & 0.026 & 2.15 &2453530.1826 &  7.70 & 16.2  & 1.0  \\   
    SL-06  & 15:30:08.77 & -43:16:04.5  &  18.6 & 6.80834 & 0.039 & 1.92 &2454615.9351 &  7.80 & 10.2  & 0.7  \\   
    SL-07  & 15:30:18.71 & -42:58:41.3  &  17.6 & 3.91403 & 0.012 & 3.24 &2453530.9068 &  9.05 & 13.4  & 1.0  \\   
    SL-08  & 15:28:56.97 & -42:55:18.2  &  18.2 & 7.08871 & 0.031 & 1.96 &2454615.9576 &  9.13 & 11.4  & 0.6  \\   
    SL-09  & 15:29:37.86 & -43:07:36.6  &  18.4 & 5.88226 & 0.035 & 1.80 &2454615.9640 &  7.53 & 12.0  & 0.6  \\   
    SL-10  & 15:29:56.55 & -42:33:30.2  &  17.9 & 5.12897 & 0.020 & 1.97 &2454528.1707 &  7.70 & 10.8  & 0.6  \\
\enddata
\end{deluxetable}

\begin{deluxetable}{lcc}
\tablewidth{0pt}
\tablecaption{Candidate Test Results\label{tab:candidates2}}
\tablehead{ 
\colhead{ID}&\colhead{Discriminant}&\colhead{Conclusion}
}
\startdata
    SL-01  & Saturated close neighbor  & Statistical False Positive  \\   
    SL-02  & High $\eta_{p}$ & EcB \\   
    SL-03  & Additional Photometric Analysis  & Blended EcB  \\   
    SL-04  & Transit depth (0.042mag) suggests EcB  & Candidate  \\   
    SL-05  & Additional Photometric Analysis  & EcB \\   
    SL-06  & Low $\alpha$ (10.2)  & Candidate \\   
    SL-07  & Lupus-TR-3  & Planet  \\   
    SL-08  & Transit correlated to $S_{j}$   & Statistical False Positive \\   
    SL-09  & Transit correlated to $S_{j}$   & Statistical False Positive  \\   
    SL-10  & Transit correlated to $S_{j}$   & Statistical False Positive \\
\enddata
\end{deluxetable}

\begin{deluxetable}{lcccccc}
\tablewidth{0pt}
\tablecaption{Monte Carlo results\label{tab:montecarlo}}
\tablehead{
\colhead{Period}&\colhead{Planet}&\colhead{Mean}&\colhead{Mean}&
\colhead{Mean}&\colhead{Detection} &\colhead{Stars} \\
\colhead{Range,}&\colhead{Radius}&\colhead{Depth}&\colhead{Duration}&
\colhead{probability}&\colhead{Efficiency}&\colhead{Probed} \\
\colhead{Distribution\tablenotemark{a}}&\colhead{($R_{J}$)}&\colhead{(mag)}&\colhead{(hr)}&
\colhead{of transit}&\colhead{($\varepsilon$)}&\colhead{($N_{pr}$)}
}
\startdata
    1-10 d, Uni    &0.8& 0.010&2.44& 0.091&0.12 & 231 \\
      &1.0& 0.016&2.49& 0.093&0.27 & 488 \\
     &1.2& 0.023&2.53& 0.095&0.43 & 762 \\
           &1.4& 0.031&2.60& 0.097&0.55 & 967 \\
    \hline
    1-3 d, Uni  &0.8& 0.010 & 1.80 &0.156 & 0.28&  697 \\
    &1.0& 0.016 & 1.81 &0.161 & 0.51& 1311 \\
         &1.2& 0.023 & 1.84 &0.164 & 0.73& 1890 \\
          &1.4& 0.031 & 1.89 &0.167 & 0.85& 2219 \\
    \hline
    3-5 d, Uni    &0.8& 0.010 &2.26 &0.095 & 0.15&  228 \\
   &1.0& 0.016 &2.30 &0.097 & 0.34&  527 \\
          &1.2& 0.022 &2.35 &0.099 & 0.54&  851 \\
          &1.4& 0.031 &2.40 &0.101 & 0.71& 1127 \\
    \hline
    5-10 d, Uni  &0.8& 0.010 &2.78 &0.063 &0.05 &46  \\
     &1.0& 0.016 &2.84 &0.064 &0.13 &141 \\
           &1.2& 0.023 &2.89 &0.066 &0.26 &275 \\
           &1.4& 0.031 &2.96 &0.067 &0.36 &392 \\
    \hline\hline
    1-10 d, Log   & 0.8 & 0.010 & 2.15& 0.120 &0.18 &416 \\
    &1.0 & 0.016 & 2.20 & 0.122 &0.37 &841 \\
         &1.2 & 0.023 & 2.23 & 0.125 &0.54 &1216\\
           &1.4 & 0.031 & 2.28 & 0.128 &0.68 &1517\\
    \hline
    1-3 d, Log   & 0.8 &  0.010 & 1.72 & 0.168 &0.28 & 759 \\
    &1.0 &  0.016 & 1.77  & 0.171 &0.54 & 1466 \\
           &1.2 &  0.023 & 1.79  & 0.174 &0.73 & 2007 \\
          &1.4 &  0.031 & 1.83  & 0.178 &0.85 & 2391 \\
    \hline
    3-5 d, Log&0.8 & 0.010 &2.24  &0.096 &0.13 &193 \\
    &1.0 & 0.015 &2.25 &0.098 &0.32 &502 \\
        &1.2 & 0.022 &2.34 &0.101 &0.53 &841 \\
          &1.4 & 0.031 &2.36 &0.102 &0.70 &1124 \\
    \hline
    5-10 d, Log &0.8 &0.010  &2.76 &0.065&0.05 &50 \\
    &1.0 &0.016  &2.79 &0.066&0.14 &152 \\
           &1.2 &0.023  &2.83 &0.068&0.26 &281 \\
           &1.4 &0.031  &2.92 &0.069&0.38 &431 \\
\enddata
\tablenotetext{a}{Uni=Uniform distribution, Log=Logarithmic distribution}
\end{deluxetable}

\begin{deluxetable}{lcccccc}
\rotate
\tablewidth{0pt}
\tablecaption{Frequency of hot Jupiters\label{tab:freq_SL}}
\tablehead{
\colhead{Period} & \colhead{Planets}  &\multicolumn{5}{c}{Planet Frequency ($f$)}\\
\colhead{Range \&} &\colhead{Detected\tablenotemark{b}} & &&&&  \\
\colhead{Distribution \tablenotemark{a}} &\colhead{($n$)}&\colhead{0.8~$R_{J}$}&\colhead{1.0~$R_{J}$}&
\colhead{1.2~$R_{J}$}&\colhead{1.4~$R_{J}$}&\colhead{mean $R_{P}$}\\
}
\startdata
1-10~d, Uni& 1  & $0.43^{+1.62}_{-0.28}\,\%$  &$<$0.62\%&$<$0.39\%&$<$0.31\%&0.16$^{+0.62}_{-0.10}\,\%$\\[3pt]
                   &[2] && [0.20$^{+0.77}_{-0.13}\,\%$]&&&[0.32$^{+0.70}_{-0.19}\,\%$]\\[3pt]
    1-3~d, Uni & 0  & $<$0.43\% &$<$0.23\%&$<$0.16\%&$<$0.14\%&$<$0.20\%\\[3pt]
    3-5~d, Uni  & 1  & $0.44^{+1.65}_{-0.28}\,\%$  &$<$0.57\%&$<$0.35\%&$<$0.27\%&0.15$^{+0.54}_{-0.10}\,\%$\\[3pt]
    5-10~d, Uni & 0  & $<$6.53\%&$<$2.13\%&$<$1.09\%&$<$0.77\%&$<$1.41\%\\[3pt]
                  &[1] && [0.71$^{+2.66}_{-0.46}\,\%$]&&&[0.47$^{+1.75}_{-0.30}\,\%$]\\[3pt]
                  \hline\hline\\
    1-10~d, Log& 1  &    0.24$^{+0.65}_{-0.18}\,\%$&$<$0.25\%&$<$0.18\%&$<$0.14\% &0.10$^{+0.27}_{-0.08}\,\%$ \\[3pt]
                        &[2] & & (0.12$^{+0.32}_{-0.09}\,\%$)  && &(0.20$^{+0.32}_{-0.14}\,\%$) \\[3pt]
    1-3~d, Log & 0  & $<$0.28\%&$<$0.15\%&$<$0.11\%&$<$0.09\%&$<$0.13\% \\[3pt]
    3-5~d, Log  & 1  & 0.52$^{+1.40}_{-0.40}\,\%$&$<$0.43\%&$<$0.25\%&$<$0.19\% & 0.15$^{+0.41}_{-0.11}\,\%$ \\[3pt]
    5-10~d, Log & 0  & $<$4.27\%&$<$1.41\%&$<$0.76\%&$<$0.50\%&$<$0.93\%  \\[3pt]
                        &[1] & & [0.66$^{+1.78}_{-0.50}\,\%$]&& &[0.44$^{+1.19}_{-0.34}\,\%$] \\[3pt]
\enddata
\tablenotetext{a}{Uni=Uniform distribution, Log= Logarithmic distribution}
\tablenotetext{b}{Statistics in square brackets are for case where
  either SL-04 or SL-06 is a planet}
\end{deluxetable}

\begin{deluxetable}{lllc}
\rotate
\tablewidth{0pt}
\tablecaption{Comparison of Planet Frequencies\label{tab:fraction}}
\tablehead{
\colhead{Survey \&} & \colhead{Method \&}&  \colhead{Assumptions} &\colhead{Planet}\\
\colhead{Reference} & \colhead{Population} & &\colhead{Frequency}\\ 
}
\startdata
    Kepler &Transit,   & $1.2<P<10$  day & $0.37\,\%$\\[2pt]
    \citep{2011arXiv1103.2541H} & ``solar-type'' & $0.7<R_{P}<1.4R_{J}$&\\[2pt]
    &dwarf stars&$n=22$&\\[2pt]
    &&&\\[3pt]

Keck Planet Search & Radial velocity,& $M_{p}>0.5~M_{J}$, & $0.65\pm0.4\%$\\[2pt]
    \citep{2008PASP..120..531C}& nearby dwarf stars &$P<5$~day&\\[2pt]
    &&&\\[3pt]

    ELODIE Planet Search & Radial velocity, & $M_{p}>0.5~M_{J}$,& $0.7\pm0.5\%$ \\[2pt]
    \citep{2005ESASP.560..833N}& nearby dwarf stars &$P<5$~day&\\[2pt]
    &&&\\[3pt]

    SWEEPS &Transit,   & $P<4.2$ day & $0.4^{+0.4}_{-0.2}\,\%$\\[2pt]
    \citep{2006Natur.443..534S} & bulge stars, & $n=16$&\\[2pt]
    &$M_{\star}>0.5~M_{\odot}$&&\\[2pt]
    &&&\\[3pt]

    OGLE-III &Transit, & n=5, $1.0<R_{P}<1.25$~$R_{J}$ &\\[2pt]
    \citep{2006AcA....56....1G} & bulge and Galactic & $\log P$ distribution,  &\\[2pt]
     & disk dwarfs stars &  $1 \le P \le 3$~day & $0.14^{+0.15}_{-0.08}\,\%$\\ [2pt]
    &  & $3 \le P \le 5$~day&$0.31^{+0.42}_{-0.18}\,\%$ \\[2pt]
    &&&\\[3pt]

    Deep MMT &Transit, &  n=0, $R=1.0~R_{J}$  & \\[2pt]
    \citep{2009ApJ...695..336H}&  Galactic disk& $\log P$ distribution & \\[2pt] 
     & dwarf stars & $0.4 \le P \le 1$~day   &  $<0.3\% $\\[2pt]
    & &$1 \le P \le 3$~day& $<0.8\%$  \\[2pt]
    &&$3 \le P \le 5$~day&  $<2.7\%$  \\[2pt]
    &&&\\[3pt]

    \textbf{SuperLupus} & \textbf{Transit,}   & \textbf{n=1,
    $\mathbf{R=1.1~R_{J}}$} & \\[2pt]
    \textbf{(this work)}& \textbf{Galactic disk} &
    \textbf{log\,\textit{P}} \textbf{distribution}\\ [2pt]
    & \textbf{dwarf stars}&\textbf{n=1 (Lupus-TR-3b)}  \\ [2pt]
    & &\textbf{$\mathbf{1 \le P \le 3}$~day}& \textbf{$<$0.15\%}\\ [2pt] 
    & &\textbf{$\mathbf{3 \le P \le 5}$~day}&\textbf{$\mathbf{0.15^{+0.41}_{-0.11}\,\%}$} \\ [2pt]
    & &\textbf{$\mathbf{ 5 \le P \le 10}$~day}&\textbf{$<$0.93 \%} \\ [2pt]
    & &\textbf{$\mathbf{ 1 \le P \le 10}$~day}&\textbf{$\mathbf{0.10^{+0.27}_{-0.08}\,\%}$} \\
\enddata
\end{deluxetable}
\end{document}